\documentclass[showpacs,preprintnumbers,10pt,twocolumn]{revtex4-1}
\usepackage{dcolumn}
\usepackage{appendix}
\usepackage{bm}
\usepackage{graphicx}
\usepackage{epstopdf}
\usepackage{amsmath,amssymb,amsthm}
\usepackage[mathscr]{eucal}
\usepackage{changes}

\begin{document}
\title{Steady state preparation of long-lived nuclear spin singlet pair at room temperature}
\author{Q. Chen}
\author{I. Schwarz}
\author{M.B. Plenio }
\affiliation{ Institut f\"{u}r Theoretische Physik, Albert-Einstein-Allee 11, Universit\"{a}t Ulm, 89069 Ulm, Germany}

\begin{abstract}
The coherent high-fidelity generation of nuclear spins in long-lived singlet states which may
find application as quantum memory or sensor represents a considerable experimental challenge.
Here we propose a dissipative scheme that achieves the preparation of pairs of nuclear spins
in long-lived singlet states by a protocol that combines the interaction between the nuclei
and a periodically reset electron spin of an NV center with local rf-control of the nuclear
spins. The final state of this protocol is independent of the initial preparation of the nuclei,
is robust to external field fluctuations and can be operated at room temperature. We show that
a high fidelity singlet pair of a $^{13}$C dimer in a nuclear bath in diamond can be generated
under realistic experimental conditions.
\end{abstract}

\maketitle
\emph{Introduction ---}
The preparation of nuclear spins in singlet states is attracting increasing attention as their
weak coupling to environmental relaxation processes makes them promising candidates for storing
nuclear hyperpolarization even beyond their relaxation time $T_1$ \cite{warren2009increasing,vasos2009long}.
Nuclear spin-singlet states offer a broad range of applications in medicine, materials science,
biology and chemistry. They are used as a resource for spectroscopic interrogation of
couplings within many-spin system \cite{pileio2007j,pileio2009extremely}, the monitoring of
protein conformational changes \cite{bornet2011long}, the probing of slow diffusion of biomolecules
\cite{ahuja2009diffusion} or as quantum memories \cite{reiserer2016robust}. However, the key strength
of nuclear singlets, namely their weak interaction with their environment due to the anti-symmetry
of the singlet state, is also their weakness, as it makes the high fidelity singlet-state preparation
and their manipulation a challenge \cite{devience2013preparation,emondts2014long}.

The nitrogen-vacancy (NV) defect center in diamond which has been studied extensively over
the past decade for precision sensing and quantum information processes (QIP) offers new perspectives
here \cite{wu2016diamond}. The NV center with its surrounding nuclei forms a natural hybrid quantum
register \cite{taminiau2014universal, waldherr2014quantum} in which electron spins are used for
fast high-fidelity control and readout, and proximal nuclear spins can be controlled and used as
memories due to their ultra-long coherence time. Furthermore, NV centers are excellent hyperpolarization
agents to polarize nearby nuclear spins at ambient condition \cite{alvarez2015local,chen2015optical,scheuer2016optically},
which gives rise to several orders of enhancement of nuclear magnetic resonance (NMR) signals.

In these room temperature applications of the NV-center, relaxation processes of the NV center
induce decoherence on both the NV and the surrounding nuclear spins, which is a major obstacle
for the high-fidelity preparation of entangled target states. However, it has been recognised
early that dissipation can also be a resource
that enables entangled state preparation \cite{plenio1999cavity,plenio2002entangled}. In recent
years theoretical protocols that design dissipative processes has been focused on the creation
of entanglement between atoms, ions and spins \cite{muschik2011dissipatively,kastoryano2011dissipative,rao2013dark,
rao2016dissipative,greiner2016purification}, the stabilisation of quantum gates \cite{ChenSP2017,bermudez2013dissipation}
and dissipative entanglement generation have been realised experimentally in atomic ensembles
\cite{krauter2011entanglement}, ion traps \cite{lin2013dissipative}, and superconducting qubits
\cite{shankar2013autonomously}.

\begin{figure}
\center
\includegraphics[width=2.8 in]{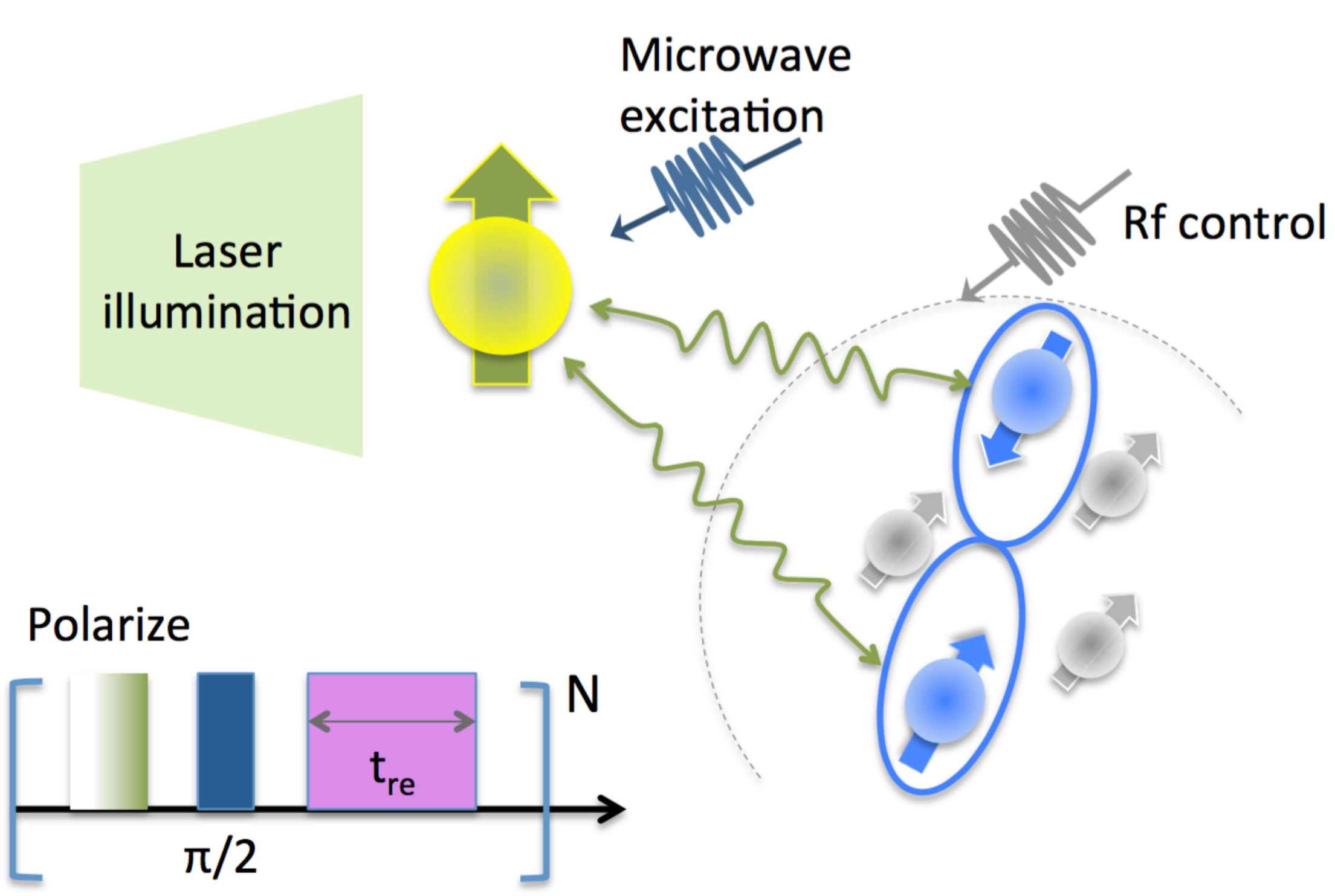}
\caption{A nuclear singlet pair (blue circles) is generated by using frequent resets of the NV
spin and local rf-control of nuclear spins. The NV spin is initialized to the $m_s=0$ state by green
laser illumination and transferred to state $|-_x\rangle = (|m_s=0\rangle - |m_s=-1\rangle)/\sqrt{2}$
by using a microwave-$\pi/2$ pulse. The electron spin is reinitialized every $t_{re}$ and MW field is
applied continuously. A unique steady state of the system is obtained by suitably chosen rf driving on the nuclear spins.}
\label{setup}
\end{figure}

Here we propose a dissipative approach to generate, at room temperature, nuclear spin-singlet
states in a spin bath in which all spins are coupled to an NV electron spin in diamond.
Our method includes two important features. First, the frequent resets of the NV
stabilises it in a particular state and provides a tunable artificial reservoir. Secondly, the coherent
local control of nuclear spins by radio-frequency fields with imbalanced detunings to the two nuclear
Larmor frequencies, ensures that the steady singlet state of the nuclei is unique. An important
merit of dissipative state preparation is its resilience to errors due to imperfect state initialization
and fluctuation of the driving fields. Our method can be applied to interacting or non-interacting nuclear
spins which have similar magnitude couplings to the NV. Additionally, high fidelity singlet pairs of a
$^{13}$C dimer in a nuclear bath in diamond can be generated with the realistic parameters. The so generated
spin singlet state exhibits a lifetime that extends well beyond the $T_1$-limit of the electron spins.

\emph{Model ---}
We consider an NV center and two nearby $^{13}$C spins with gyromagnetic ratio  $\gamma_{n}$.
Their interaction can be described by the dipole-dipole term $H_{int} = S_z \cdot \vec{A}_i\cdot\vec{I}_i$,
where $\vec{A}$ is the hyperfine vector, and non-secular terms are neglected due to the energy mismatch of
the two spins. In an external magnetic field $B_0$ (i.e, $|B_0|=100$G), the effective Lamor frequency of nuclear spin is
$\gamma_{n}|B_{eff} | =| \gamma_{n}B_0+\frac{a_{\parallel i}}{2}|$. A microwave (MW) field (Rabi frequency
$\Omega_{mw}$ and frequency $\omega_{mw}$) and a radio frequency (rf) field (Rabi frequency $\Omega_{rf}$
and frequency $\omega_{rf}$) are applied to the NV and the nuclear spins, respectively. In a suitable
interaction picture we can rewrite the effective Hamiltonian of the NV spin as $H_{NV}=\Omega_{mw}\sigma_z$
where the microwave is resonant with the $|m_s=0\rangle \leftrightarrow |m_s=-1\rangle$ transition, and
the microwave dressed states $\{|+_x\rangle=\frac{1}{\sqrt{2}}(|0\rangle+|-1\rangle), |-_x\rangle=\frac{1}{\sqrt{2}}
(|0\rangle-|-1\rangle)\}$ define $\sigma_z=\frac{1}{2}(|+_x\rangle\langle+_x|-|-_x\rangle\langle-_x|)$.
Working in a rotating frame with respect to $H_0=\Omega_{mw}\sigma_z+\sum_{i=1}^2\omega_{rf} I^z_i$ and
by using a rotating wave approximation, we find the simplified Hamiltonian (details are included in SI \cite{SI})
\begin{eqnarray}%
    H''_{tot}&=&\sum_{i=1,2}\Delta_iI^z_i+\Omega_{rf}I^{x}_i+\frac{a_{\perp i}}{4}(\sigma_+I^{-}_{i}+H.C.),
\label{Htot}
\end{eqnarray}%
with $\Delta_i=\gamma_{n}B_0+\frac{a_{\parallel i}}{2}-\omega_{rf}$ and $\omega_{rf}=\Omega_{mw}$.

We follow the basic cooling cycles
for nuclear spin polarization \cite{christ2007quantum,london2013detecting}, namely an iteration between
evolution according to Hamiltonian (\ref{Htot}) followed by reinitialization of the electron spin to
$|-_x\rangle$. The nuclei effectively \lq \lq see" a large polarisation reservoir of the periodically
reset of electron spin, and the density matrix of the system evolves according to
\begin{eqnarray}%
    \rho_n \rightarrow \cdots U_t \mathrm{Tr_e}  [U_t(\rho_n \otimes  |-_x\rangle \langle -_x|)U_t^{\dag}] \otimes |-_x\rangle\langle -_x| U_t^{\dag} \cdots \nonumber
 \label{evol}
\end{eqnarray}%
in which $U_t=\exp(-iH''_{tot} t)$ is the time evolution operator, $ Tr_e$ presents the trace over
the electron and $\rho_n$ is the density matrix of nuclear spins in the system. In order to
allow for a perturbative treatment, we consider short times between the NV resets ($t=t_{re}< 1/\sqrt{\sum_i a_{\perp i}^2}$) we can expand the
time evolution operators to second order and eliminate electronic degrees of freedom by a partial
trace. The periodic resets introduce an effective relaxation mechanism
and the effective relaxation of the NV spin is given by the life time $T_{1\rho}$ and the
reset time $t_{re}$ ($ \Gamma_N=1/T_{1\rho}+1/t_{re}$). Additionally, frequent resets with $t_{re}\ll T_{1\rho}$ ensures that
the electron spin stays close to the reset state $|-_x\rangle$ and $ \Gamma_N\approx1/t_{re}$.

By adiabatic elimination of the NV spin (see SI \cite{SI}), the reduced density operator of the
nuclear spin subsystem is governed by the master equation
\begin{eqnarray}%
    \label{LindladN}
    \frac{d}{dt}\rho_n&=&-i[H_{T},\rho_n]+\sum_{i=1,2}\mathcal{D}[M_i]\rho_n+\mathcal{D}[L]\rho_n,
    \label{D}
\end{eqnarray}%
in which $\mathcal{D}[c]\rho=c\rho c^{\dag}-\frac{1}{2}
\{c^{\dag}c,\rho\}$,  $M_i=\sqrt{\Gamma_i}I^-_i$ with $\Gamma_i$ the dephasing rate of the nuclear spins
and $L=\sum_i\alpha_iI^-_i$ with $\alpha_j=\frac{\sqrt{\Gamma_N} a_{\perp j}/4}{-\Delta_j+i \Gamma_N/2}$.
Here the effective dissipation item $\mathcal{D}[L]\rho_n$ is due to the dissipation induced by the NV
resets in combination with the interaction between the electron and nuclear spins (last term in Eq. (1))
leaving the local nuclear dynamics described by the Hamiltonian
\begin{eqnarray}%
    H_{T} =\sum_{i=1,2}\Omega_{rf}I^x_i+\Delta_{i}I^{z}_i.
\end{eqnarray}%
Without the applied rf field $\Omega_{rf}=0$ and $\Delta_{1}=\Delta_{2}$, the readily obtained master
equation represents the basic cooling scheme for nuclear spin polarization. This scheme has two decoupled
nuclear spins states, the fully polarised state $|\downarrow_1\downarrow_2\rangle$ as well as, for equally
strong coupled spins, the dark state $|S\rangle=\frac{1}{\sqrt{2}}(|\uparrow_1\downarrow_2\rangle-|\downarrow_1
\uparrow_2\rangle)$. As a result, the stationary state of the nuclear spins will be a mixture of these
two states which is neither fully polarised nor fully entangled. Our goal is the preparation of a maximally
entangled singlet state. In the following we show how local rf-control can remove the fully polarization state
of the nuclear spins from the manifold of stationary states such that the dynamics of the nuclear spins then
converges to a singlet state.

\begin{figure*}
\center
\includegraphics[width=2.5 in]{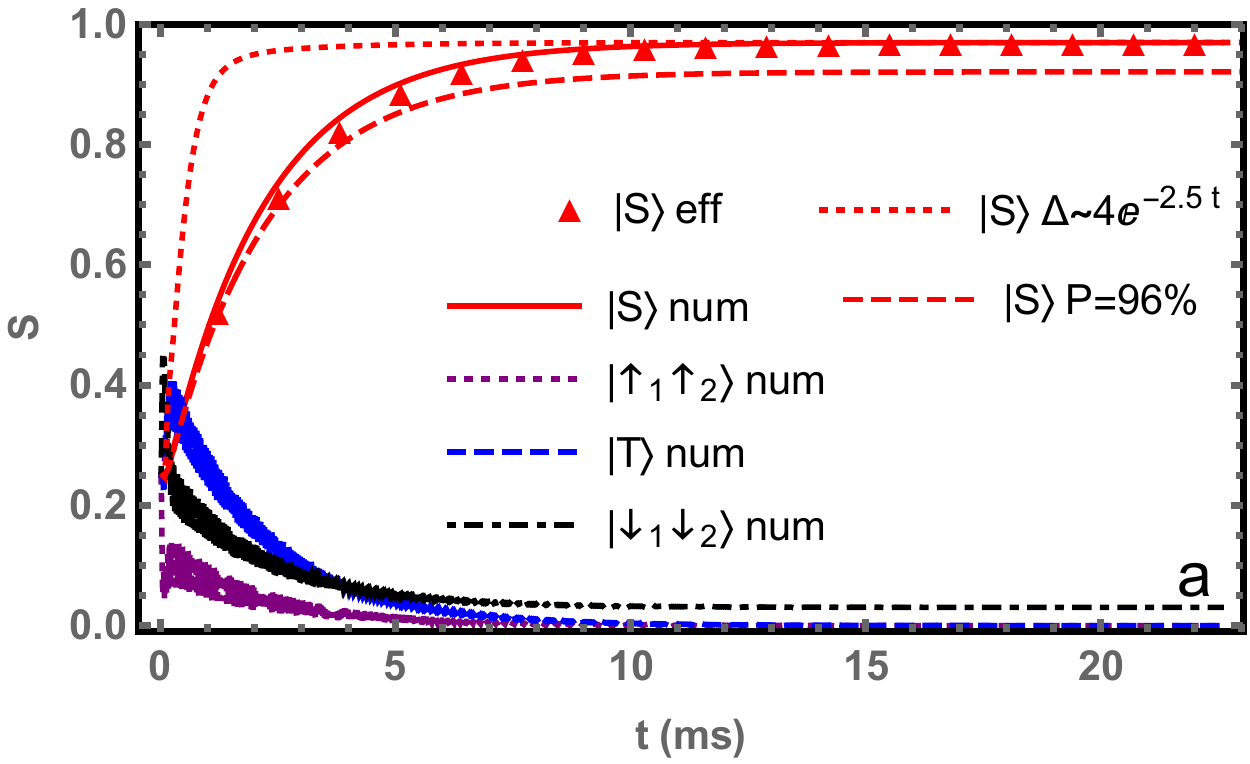}
\includegraphics[width=2.0 in]{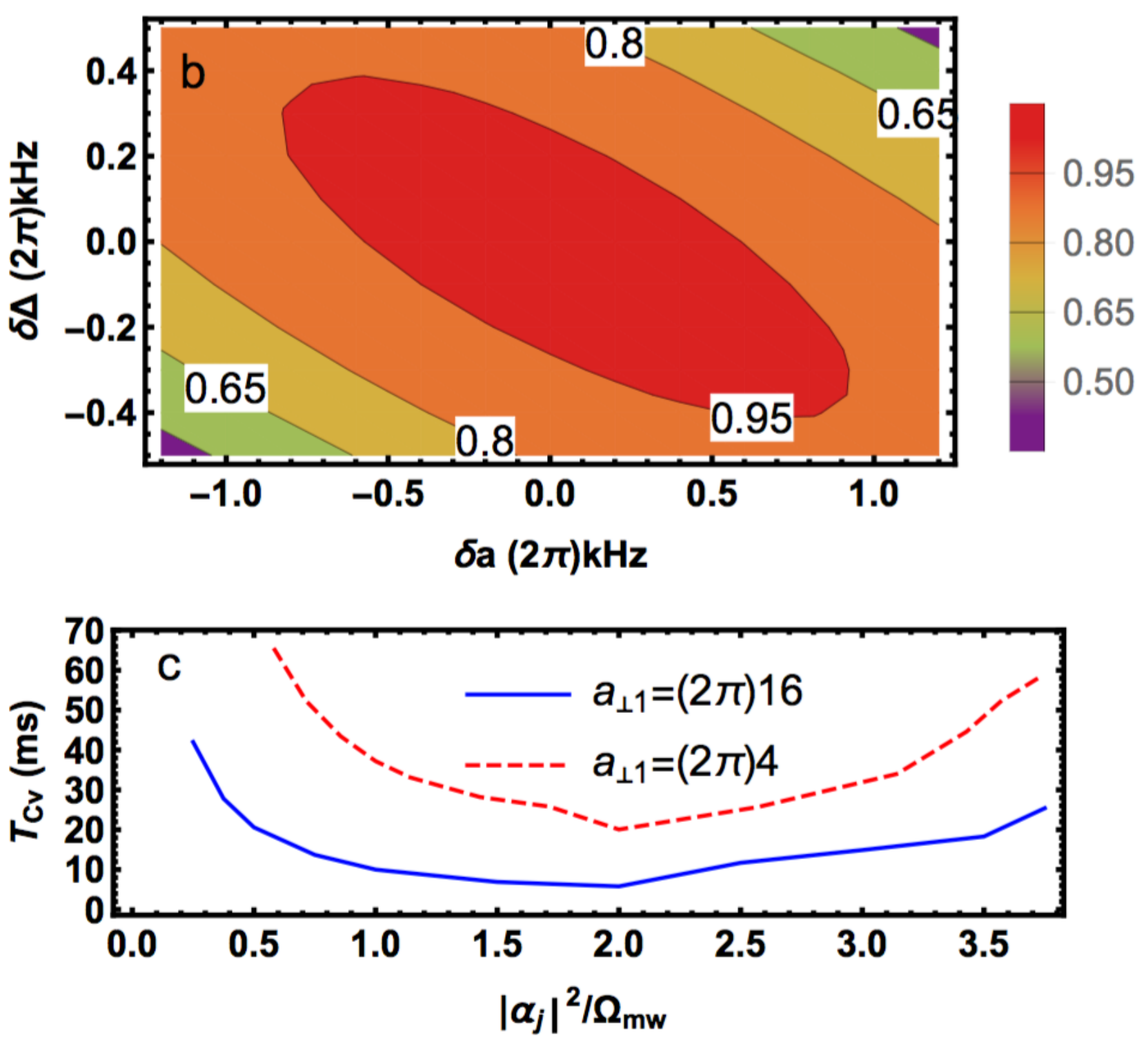}
\includegraphics[width=2.1 in]{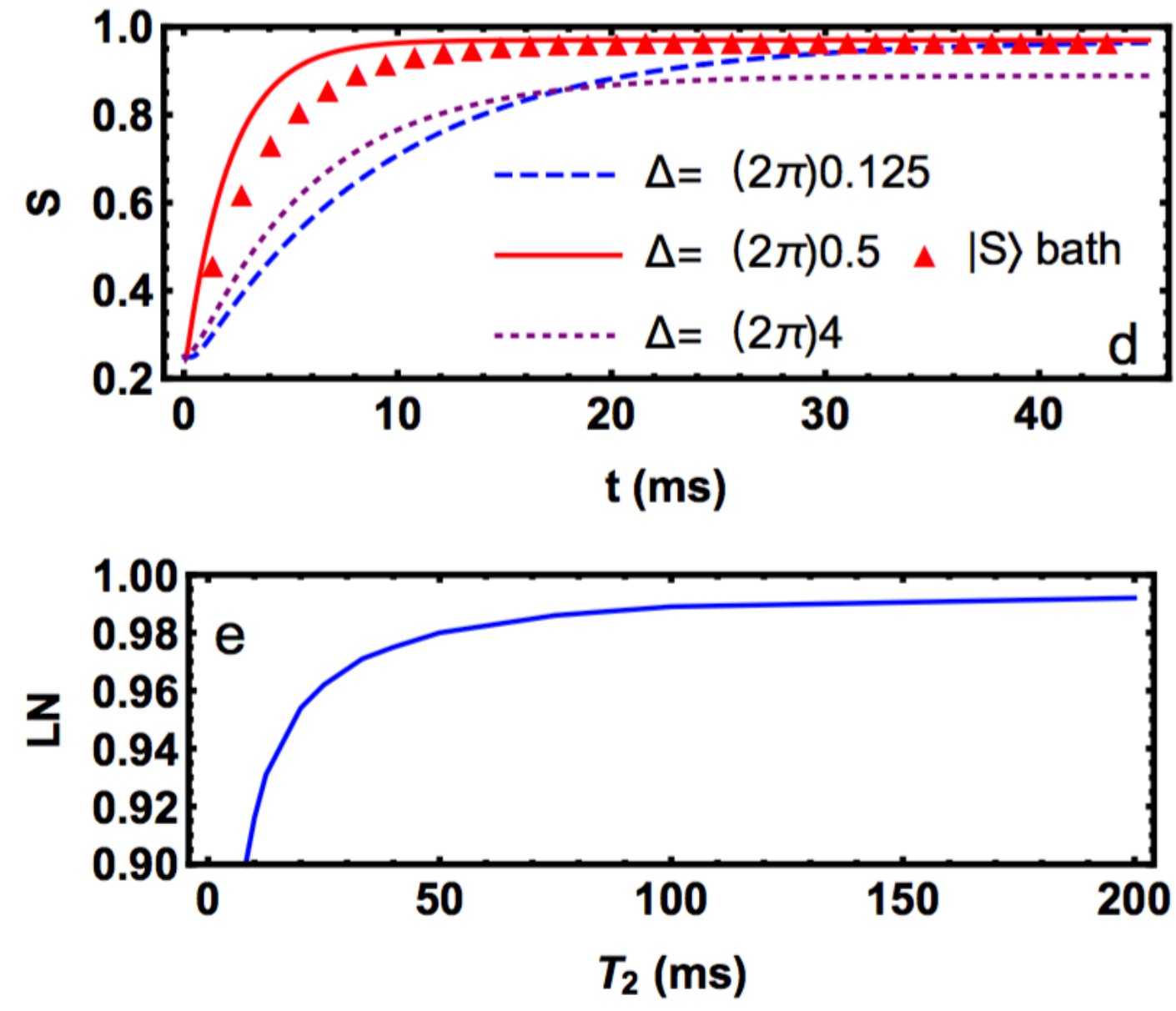}
\caption{ (a) The population evolutions of the singlet and three triplet states $\{|S\rangle=\frac{1}{\sqrt{2}}(|\uparrow_1\downarrow_2\rangle-|\downarrow_1\uparrow_2\rangle), |\uparrow_1\uparrow_2\rangle, |\downarrow_1\downarrow_2\rangle, |T\rangle=\frac{1}{\sqrt{2}}(|\uparrow_1\downarrow_2\rangle+|\downarrow_1\uparrow_2\rangle)\}$ as a function of time for a fully mixture state $\rho_n=I/4$ initially. The  NV spin is reseted every 40 $\mu$s ($T_{1\rho}=2$ ms) providing the tunable artificial reservoir. The two nuclei are coupled to the NV center (distance $\sim$1.2 nm) as $(a_{\parallel_1}, a_{\perp_1})=(2\pi)(2,16)$ kHz and $(a_{\parallel_2}, a_{\perp_2})=(2\pi)(4,16)$ kHz,  which results in $\Delta=(2\pi) 0.5$ kHz, $\Omega_{rf}=8\Delta$. Exponential detuning accelerates the convergence time (red dotted line) and imperfect reset of NV center also gives high $LN$ $\sim$0.96 of singlet state generation (red dashed line). Take the parameters as shown by the red solid line in (a) as an example to demonstrate the effect of imperfections: (b) the $LN$ is given due to imperfections of the detunings and couplings with the same $\Delta$ and evolution time $t=20$ ms. (c) The convergence time $T_{Cv}$ for achieving $LN=0.96$ vs the ratio $|\alpha|_j^2/\Omega_{mw}$ for different perpendicular couplings (Here we adjust $t_{re}$ as an example and $|\alpha|_j^2\approx a_{\perp j}^2t_{re}/4$). (d) Optimized singlet state generation of different detunings with $\Omega_{rf}=8\Delta$ and $|\alpha|_j^2/\Omega_{rf}=2$. As a comparison, we consider a small nuclear bath (red triangles), in which coupling strengths of three different nuclear spins are $(a_{\parallel_3}, a_{\perp_3})=(2\pi)(13,8)$ kHz, $(a_{\parallel_4}, a_{\perp_4})=(2\pi)(-11,3)$ kHz and $(a_{\parallel_5}, a_{\perp_5})=(2\pi)(20,4)$ kHz. And (e) with $\Delta/\Omega_{rf}=\sqrt{k_j/2}$, the optimized fidelity as a function of $T_2$ of nuclear spins.}
\label{Bfield}
\end{figure*}

\emph{Singlet pair generation for non-interacting spins ---}
For two nuclei that couple equally to the NV, i.e. $a_{\perp_1}=a_{\perp_2}$, and the choice
$\Delta_{1}=-\Delta_{2}$ for the rf-field, it is easy to see that the unique steady state of the
system is given by
\begin{eqnarray}%
    |\psi_{ss}\rangle =N_c(\sqrt{2}\Delta_{1}|\downarrow_1\downarrow_2\rangle-\Omega_{rf}|S\rangle),
\end{eqnarray}%
with the normalization coefficient $N_c=\frac{1}{\sqrt{2\Delta_{1}^2+\Omega_{rf}^2}}$. It is
an eigenstate to eigenvalue $0$ of both, the effective Hamiltonian $H_{T}$ and the Lindblad operator
$L$. Notice that when there is no detuning, i.e. $\Delta_{1}=\Delta_{2}$, the system is decomposable
and the steady state is not unique. However, a small imbalance in the detuning between the two nuclear
spins, i.e. $\Delta_{1}=-\Delta_{2}$, breaks the symmetry and leads to $|\psi_{ss}\rangle$
being the unique steady state. We use the logarithmic negativity (LN) \cite{plenio2005logarithmic,LN} to measure the entanglement, $LN(\rho_{ss})=\log_2(1+\frac{\Omega_{rf}\sqrt{2\Delta_{1}^2+\Omega_{rf}^2}}{2\Delta_{1}^2+\Omega_{rf}^2})$. Therefore, in the limit $\Omega_{rf}\gg|\Delta|$ ($\Delta=|\Delta_1-\Delta_2|/2$), one can have $LN\rightarrow 1$, which shows that the steady state of the system will achieve the singlet state $|S\rangle$ independent of the initial state.

Fig. (\ref{Bfield}a) shows the result of a numerical simulation using the original full Hamiltonian
Eq. (\ref{Htot}) and periodic NV resets. The
results which show near perfect singlet generation are well approximated by the effective master equation
Eq. (\ref{LindladN}). Even for imperfect matched detuning ($\delta \Delta=\Delta_1+\Delta_2\neq0$) and an asymmetry of the couplings $(\delta a=a_{\perp_1}-a_{\perp_2}\neq0)$ high fidelity is maintained with $\Omega_{rf}=8\Delta$ (see Fig. (\ref{Bfield}b)). Furthermore,
our simulations show that the singlet pair generation is also robust to imperfections in the reinitialization
of the NV. A reset fidelity of 96\% polarization for the electron spin (achieved by current
experimental technology \cite{waldherr2011dark}) suffices to provide a singlet state with $LN=0.96$.

Another important factor for dissipative entanglement generation is the convergence time 
$T_{Cv}$ of the scheme.  $T_{Cv}$ is limited by the effective dissipation rate $|\alpha_i|^2$ ($|\alpha_1|\approx|\alpha_2|$)
and the strength of $\Omega_{rf}$ of the local rf-control and the detuning $\Delta$ via the
ratios $\Delta/\Omega_{rf}$ and $|\alpha_i|^2/\Omega_{rf}$. Notice that $\Delta/\Omega_{rf}$ 
also controls the singlet generation fidelity and we use $\Delta/\Omega_{rf}=1/8$ to 
ensure high fidelity which in turn induces a relatively long $T_{Cv}$. For a given $\Delta$, 
we find that the optimal convergence time is achieved for $|\alpha|^2/\Omega_{rf}=2$, see 
Fig. (\ref{Bfield}c). Additionally, the convergence may be accelerated by using an adiabatic 
change of the imbalance of the detuning $\Delta$ over time. The choice $\Delta=(2\pi)4e^{-2.5t}$ 
kHz yields the dotted red line in Fig.(\ref{Bfield}a), which favours rapid approach to
the target.

The singlet state preparation can be controlled by using the weak external magnetic
field  (i.e, $|B_0|=100$G). Chosing two nuclei with similar couplings to the NV we can adjust the external magnetic field
direction to obtain the same perpendicular coupling components and a difference between parallel components.
As the detuning $\Delta$ is induced by the parallel components, choosing
the NV reset time {$t_{re}$} and Rabi frequency $\Omega_{rf}$ appropriately achieves high fidelity of singlet pair
generation and relatively short convergence time. As shown in Fig. (\ref{Bfield}d), a singlet pair is generated
with high fidelity for a large range $(2\pi)0.125$ kHz$<\Delta<(2\pi)4$ kHz. Additionally, $\Delta$ is also
tunable via a magnetic field gradient \cite{grinolds2011quantum}, which makes the adiabatic change of the
detuning imbalance $\Delta$ possible \cite{grinolds2011quantum}. Therefore it is not difficult to find two
nuclear spins matching the requirements for high fidelity generation of a nuclear singlet pair.
\begin{figure}
\center
\includegraphics[width=3 in]{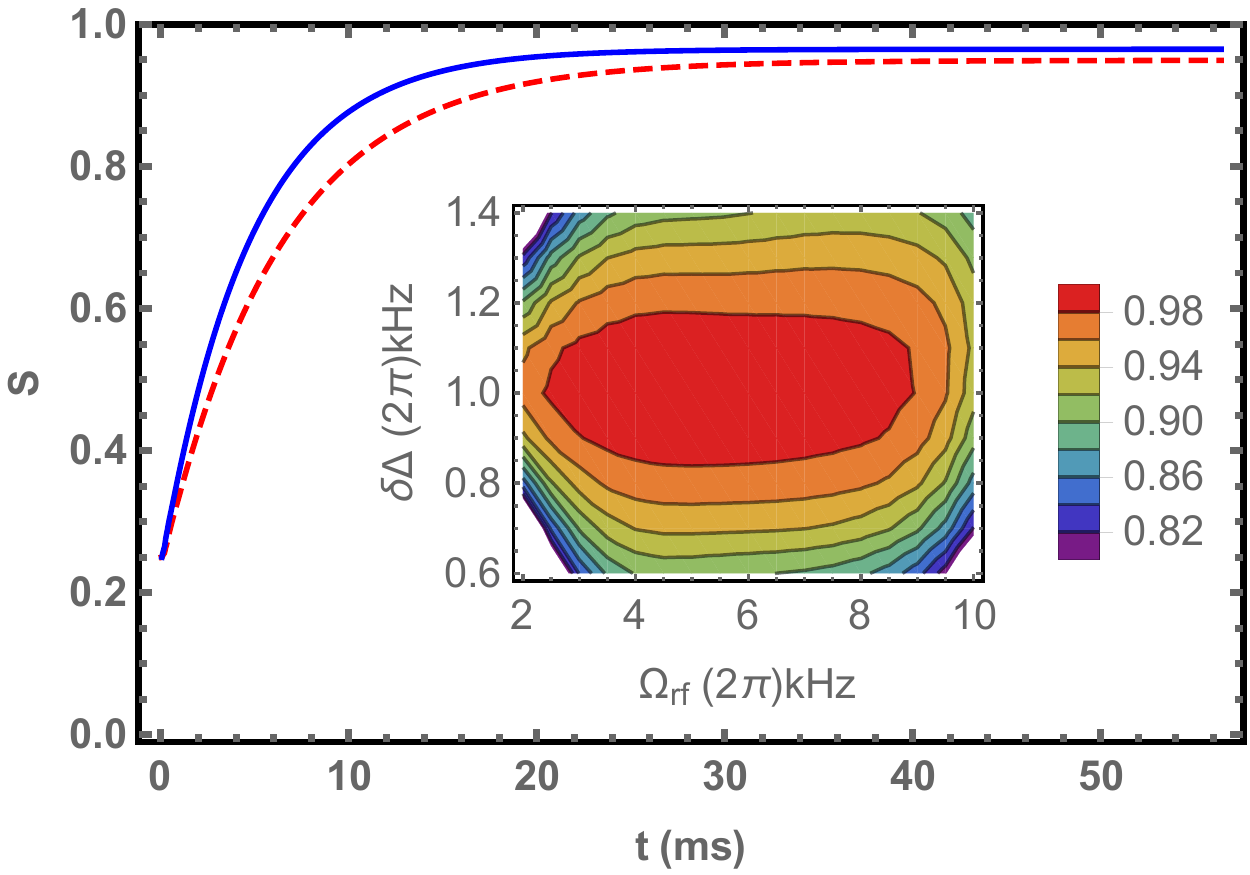}
\caption{Dissipative dynamic of the singlet state in a nuclear dimer (interacting with each other $g_{12}=(2\pi) 4.2$ kHz) coupled to the NV spin ($(a_{\parallel_1}, a_{\perp_1})=(2\pi)(-6.39,12.54)$ kHz and $(a_{\parallel_2}, a_{\perp_2})=(2\pi)(-2.77,12.67)$ kHz, which is governed by the original master equation and NV reset procedure. The initial state of nuclear pair is in a completely mixed state $\rho_n=I/4$ and NV spin is reinitialized to state  $|-_x\rangle$ every 50 $\mu$s ($T_{1\rho}=2$ ms). Adjusting the MW Rabi frequency allows us to have $\Delta_{1}=(2\pi) (-0.10)$ kHz and $\Delta_{2}=(2\pi) 1.91$ kHz,  with $\Omega_{rf}=(2\pi) 20$ kHz, and the singlet state is generated as shown by the blue line.  As a comparison, we consider a small nuclear bath (the dashed red line), in which coupling strengths of three different nuclear spins are $(a_{\parallel_3}, a_{\perp_3})=(2\pi)(-22.1,20.0)$ kHz, $(a_{\parallel_4}, a_{\perp_4})=(2\pi)(14.2,6.4)$ kHz and $(a_{\parallel_5}, a_{\perp_5})=(2\pi)(-17.8,1.2)$ kHz. The inset shows the $LN$ vs $\Omega_{rf}$ and $\delta\Delta$ with the evolution time 30 ms.  }
\label{bond}
\end{figure}

So far we have not considered the effect of environmental noise on the nuclei. In order to model a more realistic
situation, we include a small nuclear spin bath surrounding our nuclear spins pair. In Fig. (\ref{Bfield}d)
we see the impact of such a spin bath by comparing the red curve (noise free) and the red triangles (with
spin bath). Notice that the other spins are unaffected by our singlet generation protocol because $|\Delta_i-\Delta|\gg0$
and $|a_{\parallel_i}-a_{\parallel_1}|\gg0$ and $t_{re}$ is chosen such that the perturbative treatment is valid. The intrinsic decoherence of
nuclear spins can be neglected when the coherence time ($T_2>$ 500 ms \cite{maurer2012room} for $^{13}$C
spins in diamond under ambient conditions) exceeds $T_{Cv}$. For a very noisy environment, e.g.,
nuclear spins in a molecule on the NV surface the singlet fidelity will be adversely affected. One way to
eliminate the influence of intrinsic dissipation is by increasing the engineered correlated decay $\alpha_j$,
which is limited by the small coupling between the NV and nuclear spins. In order to get the steady state
of a two-qubit system, one can solve the system of
$d^2-1=15$ differential equations for the elements of the stabilized density matrix or equivalent Bloch vector.
The solution is an eighth-order polynomial in which the $LN$ can be maximized.
Therefore, one can optimize the dynamics by using the first-order perturbation in the small parameter
$k_j=\sqrt{\Gamma_j}/ \alpha_j$ , and the $LN$ is maximized by $\Delta/\Omega_{rf}=\sqrt{k_j/2}$.
We show the optimized $LN$ in Fig. (\ref{Bfield}e). Additionally, the rate of convergence can be accelerated
by starting with a large initial detuning and decrease it to the optimal value.

\emph{Singlet pair generation for interacting spins in a dimer ---}
If one intends to generate a singlet state of two interacting spins in a dimer, the dipole-dipole interaction is
not negligible and the local dynamics is given by
\begin{eqnarray}%
    H'_{T} =\sum_{i=1,2}\Omega_{rf}I^x_i+\Delta_{i}I^{z}_i+g_{12}[I^z_1I^z_2-\frac{1}{2}(I^x_1I^x_2+I^y_1I^y_2)], \nonumber
 \end{eqnarray}%
where $g_{12}=\frac{\mu_0}{4\pi}\frac{\hbar\gamma_{n}^2}{r^3_{ij}} (1-3\cos^2 \theta_{ij})$ is the coupling
strength between nuclear spins, $\theta_{ij}$ is the angle between the nuclear spin position vector $\vec{r}_{ij}$
and the magnetic field. It is easy to see that $|S\rangle$ is an eigenstate of the last term in $H'_{T}$.
Therefore, one can proceed analogously to the case of non-interacting nuclear spins and adjust the
detunings and the Rabi frequency of the rf field to ensure that the singlet state is unique. Then the
singlet state of the pair of interacting nuclear spins in a dimer is again generated as the steady state
(see the simulation in Fig. \ref{bond}). In this simulation we consider a $^{13}$C dimer ($\sim$ 1.3 nm
from NV center) in diamond, the NV position $[0, 0, 0]$nm and
NV axis is parallel to the crystal axis [111]. The magnetic field is applied along NV axis. $d_{cc}=0.154$
nm is the C-C bond length. A value of $g_{12}= (2\pi)4.2$ kHz (or 1.37 kHz) indicates the dimer is
either aligned along the direction of the external field $B$ (or tilted from the magnetic field by 109.5$^{\circ}$),
which tends to have comparable perpendicular coupling components and small imbalanced parallel components. Suppose
$g_{12}= (2\pi)4.2$ kHz, and nuclear spins positions as $[0.625, -0.624, -0.803]$ nm and $[0.536, -0.714, -0.893]$
nm, as shown in Fig. \ref{bond}, a singlet pair with high $LN$ 0.98 is generated. The scheme is robust to
the fluctuation of the detunings and Rabi frequency of rf field, see  Fig. \ref{bond}. In general, there
are many $^{13}C$ nuclear spins surrounding the NV spin. Therefore, we also performed simulations which
consider the dimer in a small nuclear bath (three additional nuclear spins coupled to the NV spin).
For 0.55\% of $^{13}C$ spins abundance, the probability of finding the dimer along the NV axis within
$1-1.5$ nm of the NV center is $\sim$2.4\% (see SI \cite{SI}).

\emph{Conclusion ---}
In summary, we propose a dissipative approach to generate long-lived singlet pair in both
non-interacting and interacting nuclear spins which are dipole coupled to the electron spin
of a nearby NV center in diamond at room temperature. The key idea is the combination
of periodic resets of the electron spin which generates tunable dissipation and coherent 
radio-frequency control of the target nuclear spins. We show that the dissipative entanglement
is generated for any initial state of the spins and is robust in the presence of
external field fluctuations and other imperfections. High fidelity nuclear singlet states
provide a resource for a host of applications.

\emph{Acknowledgements ---} This work was supported by the ERC Synergy Grant BioQ and the EU projects
DIADEMS, EQUAM AND HYPERDIAMOND as well as the DFG CRC/TR21.


\begin{thebibliography}{10}
\bibitem{LN}
$LN(\rho_{ss})= \log_2||\rho_{ss}^{PT}||_1$ is a measure of entanglement, where
  $||\cdot||_1$ denotes the trace norm and $\rho_{ss}^{PT}$ is the partial
  transpose operation appled to $\rho_{ss}$ corresponding to steady state $
  |\psi_{ss}\rangle$.
\end{thebibliography}

\end{document}